\documentclass[graybox, nosecnum]{svmult}

\usepackage{mathptmx}       % selects Times Roman as basic font
\usepackage{helvet}         % selects Helvetica as sans-serif font
\usepackage{courier}        % selects Courier as typewriter font
\usepackage{type1cm}        % activate if the above 3 fonts are
\usepackage{makeidx}         % allows index generation
\usepackage{graphicx}        % standard LaTeX graphics tool
\usepackage{multicol}        % used for the two-column index
\usepackage[bottom]{footmisc}% places footnotes at page bottom
\usepackage{hyperref}        %for hyperlinks
\usepackage{soul}            % for high-lighting of text
\def\overl#1{\mskip2mu\overline{\mskip -2mu#1}{}}
\makeindex             % used for the subject index
\usepackage{amsmath,amsbsy,amssymb}
\usepackage{isotope}
\usepackage{physics}
\begin{document}
\titlerunning{The nucleon-antinucleon interaction}
\title*{Nucleon-antinucleon interaction$^1$}

\author{Jean-Marc Richard \thanks{corresponding author}}
\institute{Jean-Marc Richard \at Institut de Physique des 2 Infinis de Lyon, IN2P3 \&\ Universit\'e de Lyon, 4 rue Enrico Fermi, Villeurbanne, France\\ \email{j-m.richard@ipnl.in2p3.fr}
}
\maketitle
\abstract{A review is presented of the antinucleon-nucleon interaction, and some related issues such as fundamental symmetries, annihilation mechanisms,  antinucleon-nucleus scattering, antiprotonic atoms and neutron-antineutron oscillations. The overall perspective is historical but the modern approaches are also presented. }
%
%\tableofcontents%
%
\footnotetext{Invited contribution to ``Handbook of Nuclear Physics'',  Springer, 2022, I. Tanihata, H. Toki and T. Kajino, Eds., Section ``Nuclear Interactions'' coordinated by R.~Machleidt}
\section{Introduction}
The positron was discovered in 1932 by Anderson \cite{1932Sci....76..238A}, and further studies confirmed that it has the expected properties, in particular, the same mass as the electron and the opposite charge.  It was thus reasonably anticipated that the proton should also possess an associated antiparticle, the antiproton, though the proton is not exactly a Dirac particle since its magnetic moment substantially deviates from the Bohr magneton, as shown by Stern in 1933 (see, e.g., \cite{Estermann1975}). 

It was also expected that the antiproton would hardly be detected with cosmic ray experiments, and a dedicated accelerator was built at Berkeley, the Bevatron (at that time, $10^9\,$eV was denoted 1\,BeV), to produce and study antiprotons. This was a remarkable technical achievement, both for the accelerator and the detector. The antiproton was discovered in 1955 by a team led by Chamberlain and Segr\`e. One year later, the charge-exchange cross-section $\bar p p\to\bar nn$ gave access to the antineutron $\bar n$, partner of the neutron $n$.  See, e.g., the Nobel lectures~\cite{foundation1998nobel}.

In successive experiments, the antiproton-proton symmetry has been checked with a greater and greater accuracy.  The charge to mass ratios $|q|/m$ turn out identical to about $10^{-10}$, and the magnetic moments $|\mu|$ to about $10^{-9}$ \cite{Zyla:2020zbs}.

\begin{figure}[ht!]
 \centering
 \includegraphics[width=.25\textwidth]{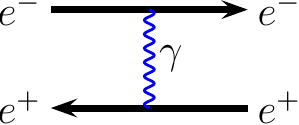}\qquad
 \includegraphics[width=.23\textwidth]{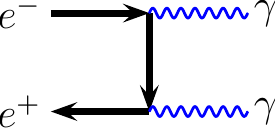}
\caption{Contributions to the elastic and annihilation $e^+e^-$ cross-sections.}
 \label{fig:e+e-}
\end{figure}

Before the Berkeley measurements, one anticipated an analogy of the $\overl N N$ interaction with the $e^+e^-$ one in QED, where the elastic interaction is mediated by photon-exchange and annihilation by fusion into two or three photons, as shown in Fig.~\ref{fig:e+e-}. 

\begin{figure}[ht!]
 \centering
 \includegraphics[width=.25\textwidth]{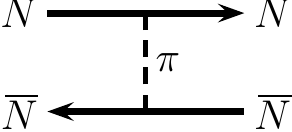}
\qquad\includegraphics[width=.28\textwidth]{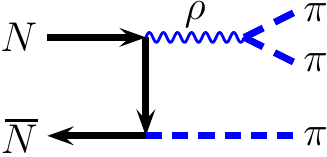}
 \caption{Simple mechanisms contributing to the $\overl N N$ interaction: one-pion exchange, and annihilation into $\rho\pi$ by baryon exchange}
 \label{fig:NNb}
\end{figure}

If one naively believes that the antiproton-proton interaction is dominated by the mechanisms  shown in Fig.~\ref{fig:NNb}, with  one-pion exchange and annihilation mediated by a very short range baryon-exchange, one predicts the following hierarchy for the annihilation ($\bar p p\to \text{mesons}$), elastic $(\bar p p\to \bar p p$) and charge-exchange $(\bar p p\to \bar n n$) cross sections
\begin{equation}
 \sigma_\text{ann}<\sigma_\text{el}<\sigma_\text{ce}~,
\end{equation}
the second inequality resulting straightforwardly from a ratio of isospin Clebsch-Gordan coefficients. Actually, the \emph{inverse} ordering was observed, with the annihilation cross section $\sigma_\text{ann}$ nearly twice larger than the elastic one, $\sigma_\text{el}$, and the charge-exchange one, $\sigma_\text{ce}$, being rather suppressed.  This pattern of the antiproton cross sections motivated a flurry of phenomenological studies which eventually forced to a drastic revision of our view of the annihilation mechanisms. 

In this review, we follow the traditional path which describes the antinucleon-nucleon interaction ($\overl N N$) as a long-range part due to meson exchanges ``supplemented'' by annihilation. In fact the main pattern of the $\overl N N$ interaction is that of a very strong annihilation with a tail of Yukawa-type contribution. Filters are required to see the latter on the top a very strong absorption. We then resume the discussion by an outline of the more modern approach, base on effective theories. For an introduction to this physics, see, e.g., \cite{Amsler:1991wbb,Dover:1992vj,Klempt:2002ap,Klempt:2005pp,Haidenbauer:2019ect}.

A word about the experimental framework. An intense program of studies of antiproton-induced has been performed, in particular at CERN and Brookhaven, using secondary beams, i.e., antiprotons just produced on a production target. Such beams had a large momentum spread and were much contaminated by other negatively-charge particles such as kaons ($K^-$) and pions ($\pi^-$). For a summary of early results, see, e.g., \cite{Flaminio:111786}.
The method of stochastic cooling has been developed at CERN~\cite{VanDerMeer:1985qh}. The main purpose was to build a high-energy collider to discover the $W^\pm$ and $Z^0$, the bosons mediating the week interactions. A low-energy branch, LEAR (Low Energy Antiproton Ring), was also set-up \cite{Koziol:2004qj}, and used to study the interaction of slow antiprotons. Stochastic cooling has also be implemented at Fermilab with high-energy $\bar p p$ collision leading in particular to the discovery of the top quark, and also medium-energy antiproton hitting a hydrogen gas target to form charmonium resonances~\cite{Garzoglio:2004kw}, as pioneered in the R704 experiment at CERN~\cite{R704:1986siy}. At CERN, the LEAR facility has been dismantled, but the ELENA decelerator gives access to very-low energy experiments~\cite{Oelert:2017mud}. For years, it has been proposed to resume experiments with antiprotons in the GeV energy range, as in the ``SuperLEAR'' project~\cite{Dalpiaz:1987qk,Hertzog:1993kt}. This will eventually be the case with the program around PANDA~\cite{Belias:2020zwx} at Darmstadt.
\section{From nucleon-nucleon to antinucleon-nucleon}
\subsection{The case of QED}
In principle, the same amplitude $\mathcal{M}(s,t)$ describes both $e^-e^-$ and $e^+e^-$ scattering, the former for $s>4\,m_e^2$ and $t<0$, and the latter for $s<0$ and $t>4\,m_e^2$. Here, $m_e$ is the electron mass, and $s$ and $t$ are two of the Mandelstam variables describing the $1+2\to 3+4$ reaction with quadri-momenta $\tilde p_i$,
\begin{equation}
 s=(\tilde p_1+\tilde p_2)^2~,\quad
 t=(\tilde p_1-\tilde p_3)^2~,\quad
 u=(\tilde p_1-\tilde p_4)^2~,
\end{equation}
which fulfill $s+t+u=m_1^2+\cdots +m_4^2$.

It is simpler, however, to compare the amplitudes $\mathcal{M}_a(s,t)$ for $e^-e^-\to e^-e^-$ and  $\mathcal{M}_b(s,t)$ for $e^+e^-\to e^+e^-$ for the \emph{same} values of $s$ and $t$. The result, sometimes referred to as the $C$-conjugation rule, is the following. If 
\begin{equation}
\begin{aligned}
 \mathcal{M}_a&{}=\left(
 \raisebox{-.2cm}{\includegraphics[width=.8cm]{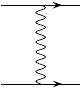}}
+\raisebox{-.2cm}{\includegraphics[width=.8cm]{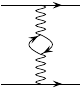}}
+\raisebox{-.2cm}{\includegraphics[width=.8cm]{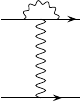}}
+\cdots\right)
+ 
\left(\raisebox{-.2cm}{\includegraphics[width=.8cm]{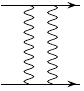}}+\raisebox{-.2cm}{\includegraphics[width=.8cm]{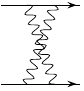}}+\cdots\right)
+\cdots \\
 &{}=\mathcal{M}_a^1+\mathcal{M}_a^2+\cdots
 \end{aligned}
\end{equation}
then $\mathcal{M}_b=-\mathcal{M}_a^1+\mathcal{M}_b^2+\cdots$, that is to say, the contributions with the exchange of an odd number of photons flip sign. 
\subsection{The case of strong interactions}
Nothing prevents from using the $C$-conjugation rule for hadrons. In particular, if 
$\mathcal{M}(pp\to pp)=\mathcal{M}_\pi+\mathcal{M}_\rho+\mathcal{M}_\omega+\cdots$ according to the mesons that are exchanged, then $\mathcal{M}(\bar p p\to\bar p p)=\mathcal{M}_\pi-\mathcal{M}_\rho-\mathcal{M}_\omega+\cdots$ 
The rule holds also for a continuum of exchanges, e.g., for two-pion exchange, $\pi^0\pi^0$ and the $C$-even part of $\pi^+\pi^-$ exchanges are unchanged when going from $pp$ to $\bar p p$, while the $C$-odd part of $\pi^+\pi^-$ flips sign. 

However, it is convenient to relate amplitudes with the same isospin, for instance, $pp\to pp$ for $NN$ and $\bar n p\to \bar n p$ for $\overl N N$. This is achieved with the $G$-parity rule of Fermi and Yang \cite{Fermi:1959sa}: if in isospin $I$
\begin{equation}
 \mathcal{M}^I(NN)=\mathcal{M}^I_\pi+\mathcal{M}^I_{\pi\pi}+\mathcal{M}^I_\omega+\cdots,
\end{equation}
 according to the mesons that are exchanged ($\rho$ is included here in $\mathcal{M}^I_{\pi\pi}$), then
\begin{equation}
 \mathcal{M}^I(\overl N N)=-\mathcal{M}^I_\pi+\mathcal{M}^I_{\pi\pi}-\mathcal{M}^I_\omega+\cdots
\end{equation}
This was the basis of nearly 50 years of phenomenological studies.
%
%\boldmath
\subsection{Consequences of the $G$-parity rule}
%\unboldmath
%
First a remark on the isospin structure. The charge-exchange cross section being governed by
\begin{equation}
 \mathcal{M}_\text{ce}\propto \mathcal{M}^0(\overl N N)-\mathcal{M}^1(\overl N N)~,
\end{equation}
its suppression results from a cancellation  between the $I=0$ and $I=1$ amplitudes. This turns out to be a very severe constraint. 

The second observation deals with the average properties of the long-range interaction. The $NN$ potential is weakly attractive, as seen from the shallow binding of the deuteron, and the absence of dineutron. In the conventional picture of  meson exchanges, this is understood by some cancellation between the attractive contribution of $f_0$ exchange\footnote{Better known in the nuclear physics community as $\epsilon$ or $\sigma$ exchange, or as the scalar-isoscalar part of two-pion exchange}, and repulsive contributions such as $\omega$ exchange. Once the $G$-parity rule is applied, one gets a coherent attraction in the $\overl N N$ case. Of course, the risk is that models with large coupling constants, but properly tuned by cancellations to reproduce the $NN$ data, would give an unrealistic $\overl N N$ attraction. 

Already in \cite{Fermi:1959sa}, and later in the ``bootstrap'' era \cite{Chew:107781}, this attraction suggested a composite picture of mesons as $\overl N N$ bound states. Among the many difficulties in this approach, one can mention the breaking of ``exchange degeneracy'': for instance the $\omega$  meson with isospin $I=0$ and the $\rho$ meson with $I=1$, both with spin-parity $J^P=1^-$, have nearly the same mass, while the meson-exchange part of the $\overl N N$ interaction is more attractive for $I=0$ than for $I=1$~\cite{Ball:1965sa}. 

Two decades after Fermi and Yang \cite{Fermi:1959sa}, Shapiro and his collaborators~\cite{Shapiro:1978wi}, and others \cite{Buck:1977rt}, came back on this idea, and proposed that the $\overl NN$ bound states and resonances are associated with new kind of mesons, preferentially coupled to $\overl NN$, named ``quasi-nuclear'' states or ``baryonia''. We return to this question in the section devoted to hadron spectroscopy. 

Another consequence of the $G$-parity rule is a change of the spin dependence of the interaction. In the $NN$ case, the most salient feature is the presence of a strong spin-orbit component, at work in nucleon scattering and in the spectroscopy of nuclear levels. For $\overl NN$, the spin-orbit component is moderate, but a very strong coherence is observed in the tensor component, especially for $I=0$. In nuclear physics, it is well known that the tensor potential is crucial to achieve the binding of the deuteron, but the fraction of $^{3}D_1$ admixture into the dominant $^{3}S_1$ remains small, of the order of 5\,\%.\footnote{In the limit of very strong tensor forces, this percentage can reach~2/3.}\@ In the case of $\overl N N$, there are cancellations among the various meson exchanges contributions to the spin-orbit potential, but coherences for the tensor component, especially for $I=0$~\cite{Dover:1978br}.  
\section{Optical models}
Once the long-range $\overl N N$ interaction is derived from the $NN$ one by the $G$-parity rule, it has to be supplemented empirically by short-range terms. 
The  elastic ($\bar p p\to \bar p p$), charge-exchange ($\bar p p\to \bar n n$) and annihilation ($\bar p p\to \text{mesons}$) integrated cross-sections have been analyzed by Ball and Chew \cite{PhysRev.109.1385}, L\'evy \cite{PhysRevLett.5.380}, \dots, who concluded that one needs a strong absorption even in the partial waves with angular momentum $\ell>0$. This was confirmed in explicit fits with optical potentials~\cite{1958NCim....8..485G,Bryan:1968ns}.

One may wonder: why an optical potential? Because this is a valuable tool to study antiprotonic atoms, antiproton-nucleus and antinucleus-nucleus interaction, etc. Some of the applications will be outlined in the next sections. 
If, for instance, one studies the strangeness-exchange reaction $\bar p p\to \overl\Lambda\Lambda$, one should in principle setup a cumbersome system of coupled channels $\bar p p\leftrightarrow \text{mesons}\leftrightarrow\overl\Lambda\Lambda$, in which the mechanism of $K$, $K^*$ exchange, or internal conversion $\pi\pi\to \overl K K$ could be somewhat lost. Instead, an optical model provides the adequate distorted waves for the initial and final states. 

Anyhow, further optical potentials were elaborated in the same spirit as \cite{Bryan:1968ns}, namely: start from a meson-exchange model of the long-range nucleon-nucleon interaction, apply the $G$-parity rule, and replace its short-range part by an empirical complex core whose parameters are adjusted to fit the $\overl N N$ data. One may cite the Dover-Richard \cite{PhysRevC.21.1466} or  Kohno-Weise \cite{Kohno:1986fk} potentials, with always the same conclusion: one needs a very strong absorption up to at least $0.8\,$fm. In a baryon-exchange picture, this implies very wide form factors; this means it becomes more appropriate to speak of the size rather than of the range of annihilation. See, also, \cite{Mull:1994gz}. 
\section{Spin observables}
There has been early measurements of the polarization (or analyzing power) in the elastic reaction. In the LEAR era, it could have been envisaged a thorough investigation of the spin observables, but this was not approved by the CERN committee in charge, in view of the very packed program of experiments. The aim was twofold: probe our understanding of the $\overl NN$ interaction, and, possibly, build a set-up for polarizing antiprotons, by filtering or transfer.  For a summary of the available data, see, e.g., \cite{Klempt:2002ap,Haidenbauer:2019ect}.

\begin{figure}[ht!]
 \centering
 \includegraphics[width=.5\textwidth]{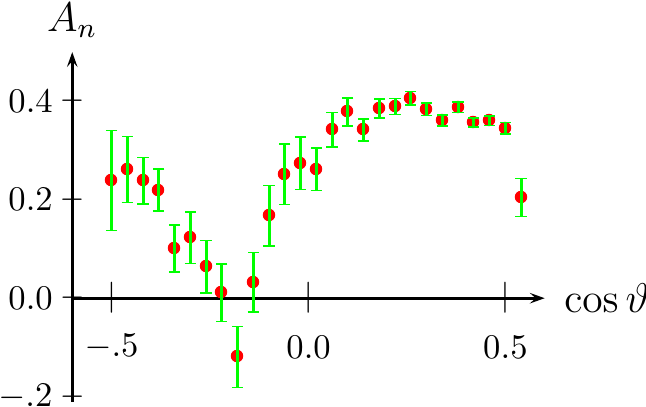}
 % An679PS172-fig1.pdf: 186x117 px, 72dpi, 6.56x4.13 cm, bb=0 0 186 117
 \caption{Analyzing power of $\bar p p\to\bar p p$ at $p_\text{lab}=679\,$MeV/$c$~\cite{Kunne:1988tk}.}
 \label{fig:An-679}
\end{figure}

Figure~\ref{fig:An-679} shows the analyzing power of elastic $\bar p p$ scattering at momentum $p_\text{lab}=679\,$MeV/$c$~\cite{Kunne:1988tk}. The moderate values can be understood to be due to a moderate spin-orbit component, or to a strong tensor component acting beyond first order, or to a combination of both. 

As for the phenomenology of spin observables, it has been noticed that  they are sensitive to the high partial waves and thus to the meson-exchange tail of the interaction. Simulations have been attempted with a spin- and isospin-independent complex core and a long-range part given by the $G$-parity rule \cite{Joseph:1981zv,Dover:1981pp}. The polarization (or analyzing power) is rather moderate, but some rank-2 observables sensitive to the tensor interaction are rather pronounced. This implies that experiments should be performed with two simultaneous spin measurements, and beam or target polarized longitudinally rather than transversally. A striking prediction is that a charge-exchange reaction on a longitudinally-polarized proton target will produce polarized antineutrons. 

\begin{figure}[ht!]
 \centerline{
 \includegraphics[width=.51\textwidth]{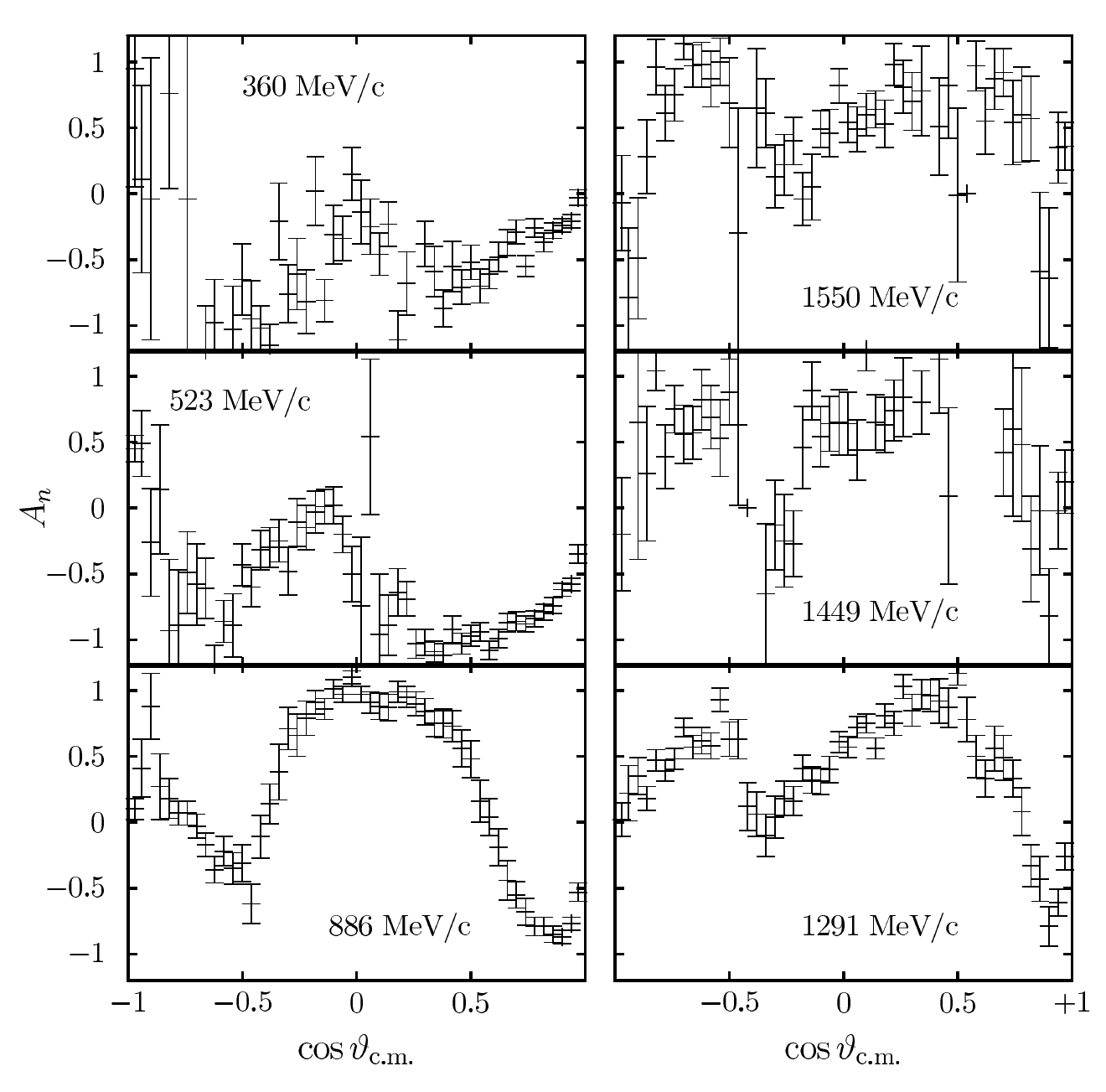}
 \quad
  \includegraphics[width=.51\textwidth]{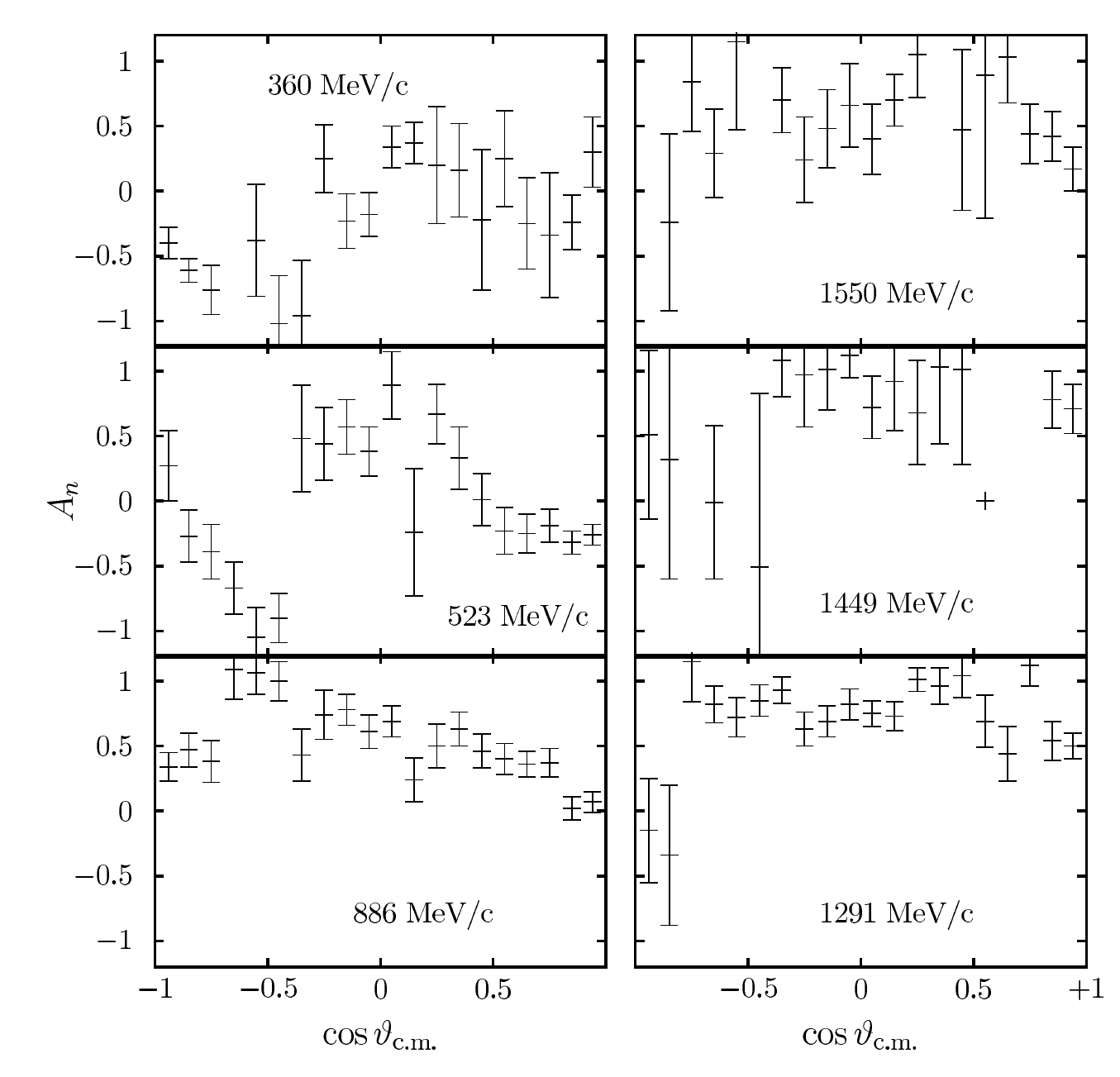}}
 % Polpipi.pdf: 380x375 px, 72dpi, 13.41x13.23 cm, bb=0 0 380 375
 \caption{Asymmetry in the reactions $\bar pp\to\pi\pi$ (left) and  $\bar pp\to KK$ (right), as measured by the PS172 collaboration~\cite{Hasan:1992sm}.}
 \label{fig:pipi}
\end{figure}

Interesting spin effects are not restricted to $\overl NN\to\overl N N$. They are also at work in other $\overl N$-induced reactions such as $\bar p p\to \pi^+\pi^-$, $K^+K^-$ or $\overl \Lambda\Lambda$. 
The latest measurements of annihilation into two pseudoscalars, $\pi\pi$ or $KK$, have been done by the collaboration PS172 with a polarized target~\cite{Hasan:1992sm}. The analyzing power (or asymmetry, somewhat improperly called polarization) turns out extremal ($|A_n|\sim 1$) in some wide ranges of energy and angles, as seen in Fig.~\ref{fig:pipi}. This means that one of the transversity amplitudes dominates. The pattern observed in Fig.~\ref{fig:pipi} has been explained from different viewpoints, such as initial~\cite{Elchikh:1993sn} or final state \cite{Takeuchi:1992si}
interaction. 

The $\bar p p\to\overl\Lambda\Lambda$ (and also some other hyperon-antihyperon final states) reaction has been measured at LEAR by the PS185 collaboration. Even without polarized target, interesting results can be obtained \cite{Barnes:1996si}, as the weak decay of $\Lambda$ or $\overl\Lambda$ informs about its spin. The most striking result is the suppression of the spin-singlet fraction. A complete reconstruction was possible with the data taken using a polarized proton target~\cite{Paschke:2006za}. The detailed measurement of the $\bar p p\to\overl\Lambda\Lambda$  has reactivated the studies about the constraints among the various spin observables. Very often, when two observables $X$ and $Y$ which belong to the interval  $[-1,+1]$, the set $\{X,Y\}$ is restricted to a subset of the square $[-1,+1]^2$, such as the disk $X^2+Y^2\le 1$ or the triangle $Y-2\,|X|+1\le0$. For triples of observables, a variety of subdomains of the cube $[-1,+1]^3$ are obtained~\cite{2009PhR...470....1A}. Checking such constraints is a prerequisite for any amplitude analysis. 

\section{$\overl NN$ interaction in effective theories}
\begin{figure}[ht!]
 \centerline{
 \includegraphics[width=.12\textwidth,height=.09\textwidth]{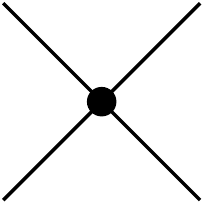}\qquad
 \includegraphics[width=.12\textwidth,height=.09\textwidth]{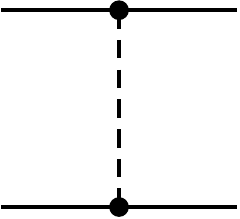}\hfill}
 \vskip .3cm
 \centerline{\includegraphics[width=.12\textwidth,height=.09\textwidth]{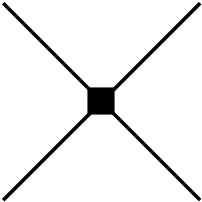}\qquad
 \includegraphics[width=.12\textwidth,height=.09\textwidth]{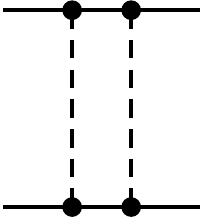}\qquad
 \includegraphics[width=.12\textwidth,height=.09\textwidth]{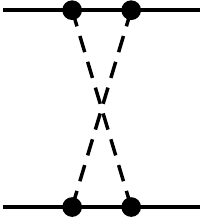}\qquad
 \includegraphics[width=.12\textwidth,height=.09\textwidth]{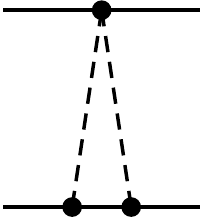}\qquad
 \includegraphics[width=.12\textwidth,height=.09\textwidth]{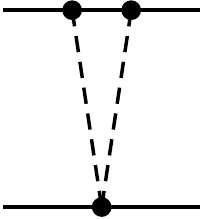}\qquad
 \includegraphics[width=.12\textwidth,height=.09\textwidth]{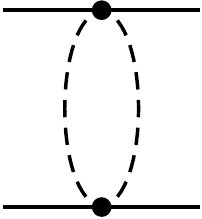}\hfill}
 \caption{Some contributions to the $\overl N N$ interaction in chiral effective field theory. The disk and square contact terms differ by the number of powers of the momenta. }
 \label{fig:chi}
\end{figure}

Chiral effective theory has been proposed by Weinberg for the study of the nucleon-nucleon interaction and other hadronic systems at low-energy. It has become very fashionable. The $NN$ potentials based on this approach have now become the ultimate standard in the field. 
For an introduction, see, e.g., \cite{vanKolck:2021rqu}, and for the application to $\bar N N$, \cite{Dai:2017ont} and refs.\ there. In this approach, the $\overl N N$ interaction consists of one-pion-exchange and a series of contact terms with an increasing power of the incoming and outgoing momenta. A sample of diagrams picturing the first contributions is shown in Fig.~\ref{fig:chi}. For more details, see~\cite{Dai:2017ont}. 
The aim is to provide a more consistent and systematic approach. However, the low-energy constants are tuned to fit a set of  $\bar NN$ phase-shifts that are not free from ambiguities. This strategy could perhaps be improved. Remember that early $NN$ potentials were fitted to reproduce some phase-shifts (see, e.g., \cite{DeTourreil:1973uj}), while more recent models are tuned directly to the $NN$ observables (see, e.g., \cite{Lacombe:1980dr}).

\section{Annihilation mechanisms}
\subsection{General considerations}
Several important questions are related to annihilation:
\begin{enumerate}
 \item Can one account for the overall strength of the absorptive component of the $\overl N N$ interaction ?
 \item Can one understand the main patterns of the observed branching ratios $\bar p p\to \pi\pi$, $\bar K K$, $\pi\pi\pi$, \dots at rest and in flight? What is the role of final-state interaction?
 \item Is annihilation an adequate doorway to study the spectroscopy of the light mesons and identify some light exotics?
 \item Is the optical model well suited to account for annihilation?
 \item Can one understand annihilation in terms of the quark content of the nucleon and antinucleon?
\end{enumerate}
Anyhow, annihilation is a fascinating and long-debated issue, and there is no consensus at present whether annihilation shall be described at the hadronic level or at the quark level. 
\subsection{Baryon exchange}
\begin{figure}[h!]
 \centering
 \includegraphics[scale=.82]{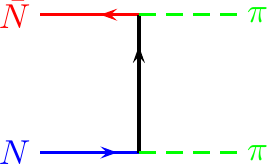}
 \quad
\raisebox{-2.5pt}{\includegraphics[scale=.78]{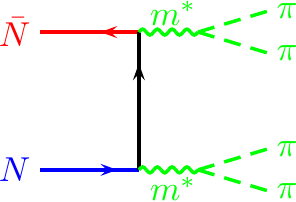}}
   \quad
  \includegraphics[scale=.78]{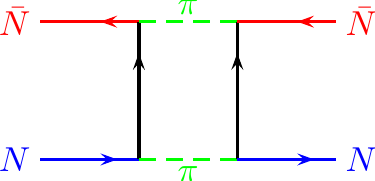}
    \quad
  \includegraphics[scale=.78]{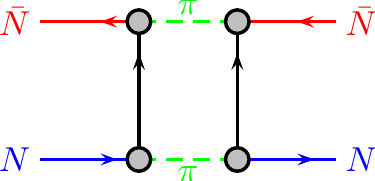}
 \caption{Some annihilation diagrams driven by baryon exchange: production of two pions, production of several pions though meson resonances, iteration of two-pion production without and with form factors.}
 \label{fig:bar-exch}
\end{figure}

As reminded in the introduction, the analogy with the $e^+e^-$ annihilation in QED suggests a mechanism of baryon exchange, with some examples in Fig.~\ref{fig:bar-exch}. Also shown is an example of iteration, that contributes to the $\overl N N$ amplitude. The large mass of baryons implies a range of about 0.1\,fm, but some form factor corrections are in order, in which case it is more appropriate to talk about the size rather than the range of the annihilation. Note that if the nucleon or nucleon resonance is replaced by a hyperon, then some strange mesons can be produced. 
The appealing aspect is that one can build $\overl N N$ potential in which both the elastic and absorptive parts are described in terms of hadrons, with the same coupling constants, see, e.g.~\cite{Hippchen:1991rr}. However, all the diagrams with baryons in the mass range 1-3\,GeV have about the same range, and it is not clear how the series of diagrams converges.
\subsection{Annihilation viewed in terms of quarks}
\begin{figure}[ht!]
 \centering
 \includegraphics[width=.25\textwidth]{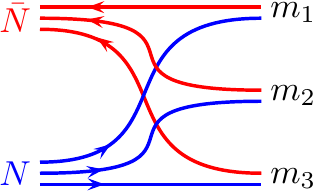}
 % diag-quarks-fig1.pdf: 128x84 px, 72dpi, 4.52x2.96 cm, bb=0 0 128 84
 \caption{Rearrangement of the three quarks of a nucleon and the three antiquarks of an antinucleon into three mesons.}
 \label{fig:rearr}
\end{figure}

Already at the beginning of the quark model, or, say, of the SU(3) flavor symmetry, a systematics of the branching ratios was attempted (without an estimate of the overall annihilation cross-section)~\cite{Rubinstein:1966zza}. The mechanism corresponds to the rearrangement, as schematically shown in Fig.~\ref{fig:rearr}.
It was revisited in the~80s, in particular by Green and Niskanen~\cite{Green:1985yy} and by Pirner et al.~\cite{Ihle:1988mp}, who demonstrated that it gives the right order of magnitude for the strength of annihilation. 

The validity of this approach was much debated. First it was argued that the annihilation has to be short-ranged, on the basis of very general properties of scattering amplitudes~\cite{PhysRev.124.614}. But, again, this is not a question of range, but a problem of size, as strictly speaking, rearrangement is not ``annihilation''. Once the diagram of Fig.~\ref{fig:rearr} is estimated with harmonic oscillator wave functions for mesons and baryons, or, more precisely, its iteration $\overl N N\to\text{mesons}\to\overl NN$ can be estimated, and the result is a separable interaction~\cite{Green:1985yy,Ihle:1988mp} with the form factors directly related to the size of the hadrons. 

\begin{figure}[ht!]
 \centering
 \includegraphics[width=.99\textwidth]{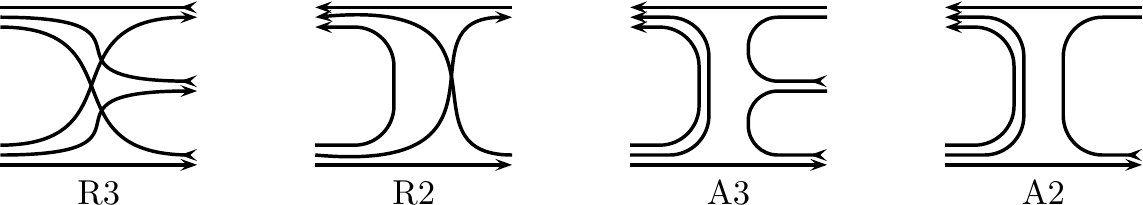}
 \caption{Some diagrams contributing to the $\overl N N $ annihilation.}
 \label{fig:annih:diags}
\end{figure}

Another concern is related to  the belief that planar diagrams should dominate. In the terminology spelled out in Fig.~\ref{fig:annih:diags}, the planar diagrams are A2 or A3. 
The dominance of planar diagrams has been revisited by Pirner~\cite{Pirner:1988mn}, who concluded that the rearrangement diagrams is not suppressed. The respective role of the A2, A3, R2 and R3 diagrams was also discussed phenomenologically, see, e.g., \cite{Dover:1986jh,Maruyama:1987tx,Dover:1992vj}. Clearly, R3 alone cannot produce kaons, unless some rescattering such as $\pi\pi\to\bar K K$ is introduced. On the other hand, A2 and A3 tend to produce too many kaons, and thus require the introduction of an empirical ``strangeness suppression factor''. 
\subsection{Phenomenology of annihilation}
Many data have been accumulated over the years, in particular at Brookhaven and CERN. The measurements at rest are listed together with the pressure in the target, from which one can infer the percentage of S-wave annihilation, as seen in the section on antiprotonic atoms. 

One should stress that probing an annihilation mechanism is not an easy task. Each branching ratio is typically of the form
\begin{equation}
 B(\overl N N\to m_1\,m_2\,\ldots)=
 \mathcal{P}(\overl N N)\times \text{PS}\times 
 |\mathcal{M}(\overl N N\to m_1\,m_2\,\ldots)|^2~,
\end{equation}
involving the probability $\mathcal{P}$ to find $\overl NN$ in the appropriate partial wave, the phase-space factor PS, and the square of the amplitude. In protonium (for annihilation at rest), and to a lesser extent in flight, there are dramatic differences among the various S-wave or P-wave  probabilities $\mathcal{P}$ corresponding to different spin or isospin. Predicting the $\mathcal{P}$ factors would require a reliable model of the $\overl N N$ interaction probed with spin-observables that have never been measured. For instance, an observation that $B(\rho\phi)\ll B(\omega\phi)$ would indicate either that the short-range protonium wave function has more $I=0$ than $I=1$, or that the quark diagram interfere constructively for $\omega\phi$ and destructively for $\rho\phi$. 
\section{$\overl{N}N$ interaction and hadron spectroscopy}
The physics of antiprotons has always been closely related to the hadron spectroscopy. Even before the era of stochastic cooling, a lot data have been collected on annihilation at rest, and several light meson resonances have been identified thanks to $\overl N N$. This search has been resumed at LEAR, in particular with the Asterix, Obelix and Cristal-Barrel experiments with the aim to detect new kinds of light mesonic resonances, for instance $q\bar q g$ hybrids. For a review of light-meson production at LEAR, see, e.g., \cite{Klempt:2005pp,Amsler:2019ytk}.

Historically, the physics of low-energy antiprotons has been developed in the 80s to study baryonium.  The name ``baryonium'' denotes mesons that are preferentially coupled to baryon-antibaryon channels. The baryonium was predicted by Rosner~\cite{Rosner:1968si}, on the basis of duality arguments (for a review, see, e.g., \cite{Phillips:1974mr}): schematically, a reaction $a+b\to c+d$ can be described either as a sum of $s$-channel resonances or $t$-channel exchanges; then, a coherent dynamics in the $t$-channel implies the existence of $s$-channel resonances; for $\overl N N$, the meson-exchanges in the $t$-channel are dual of $s$-channel $\overl NN$ resonances. Some indications were found in the 70s, as reviewed by Montanet in \cite{Montanet:1980te}, and these discoveries have motivated the construction of the low-energy facility LEAR at CERN. Unfortunately, none of the peaks discovered in antiproton-induced reactions were confirmed at LEAR. On the other hand, baryon-antibaryon pairs are sometimes observed in the decay of heavy quarkonia of flavored mesons. See, e.g., the review on non-$q\bar q$ mesons by Amsler and Hanhart in~\cite{Zyla:2020zbs}.

\begin{figure}[ht!]
 \centering
\raisebox{.7cm}{\includegraphics[width=.20\textwidth]{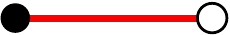}}
 \qquad
 \includegraphics[width=.20\textwidth]{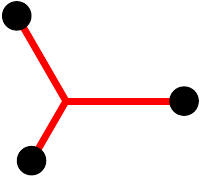}
 \qquad
 \includegraphics[width=.25\textwidth]{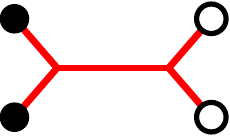}
 \caption{String picture of mesons, baryons and tetraquarks (baryonia in the light sector)}
 \label{fig:bary}
\end{figure}

On the theory side, there has been several approaches to baryonium, anticipating the complementarity and emulation between the tetraquark models and the molecular picture of exotics. The baryonium has been modeled as a color-$\bar 3$ diquark and a color-$3$ antidiquark (without any strict derivation of such clustering). Even more speculative is the so-called mock-baryonium with a color-$6$ diquark and a color-$\bar 6$ antidiquark.  In the string dynamics, one can view mesons, baryons and baryonia as successive stages of the construction, as pictured in Fig.~\ref{fig:bary}. For its link to QCD, see e.g., \cite{Montanet:1980te,Rossi:2016szw}.

Another approach is based on the $\overl N N$ interaction. Unlike some speculations in the 60s, tentatively associating $\overl N N$ states with ordinary mesons, Shapiro and his followers associated such $\overl N N$  states with new kind of mesons, the baryonia~\cite{Buck:1977rt,Shapiro:1978wi}. This was named ``quasi-nuclear'' picture, but it became ``molecular''. In the above references, the $\overl N N$ spectrum was first calculated using the real part of the interaction, and then the effect of annihilation was discussed in a rather empirical (and optimistic) manner. More serious calculations, using the whole optical potential, have shown that most of the states are washed out by annihilation \cite{Myhrer:1976ka,Dalkarov:1977xp}, while a few states might survive~\cite{Wycech:2015qra}.

% \begin{figure}[ht!]
%  \centering
%  \includegraphics[width=.5\textwidth]{FigsExt/Diff-pp.png}
%  % Diff-pp.png: 600x600 px, 72dpi, 21.17x21.17 cm, bb=0 0 600 600
%  \caption{Differential cross section, as measured by}
%  \label{fig:diffpp}
% \end{figure}
%
\section{Antinucleon-nucleus interaction}
\subsection{Elastic scattering}\label{subse:pbar-A}
\begin{figure}[ht!]
 \centerline{
\includegraphics[width=.35\textwidth]{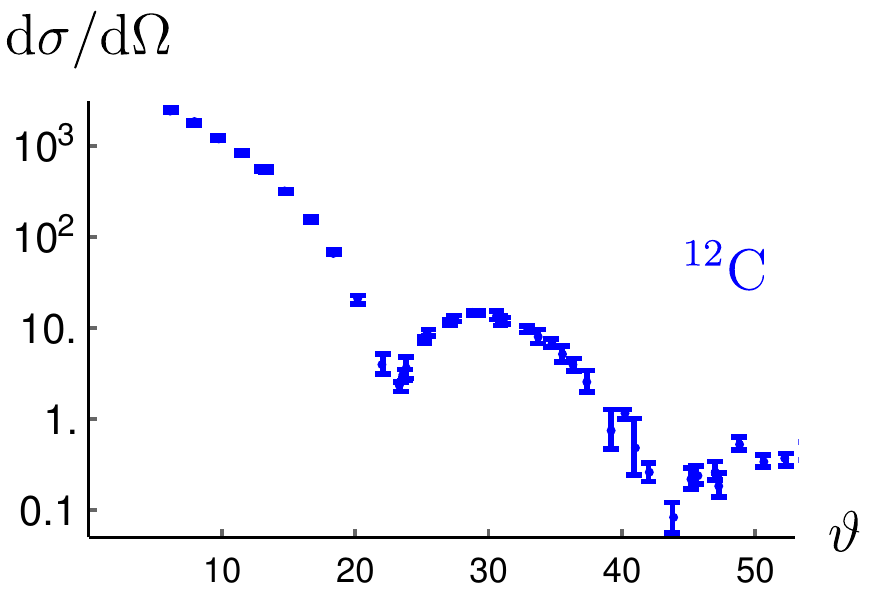}
\quad \includegraphics[width=.35\textwidth]{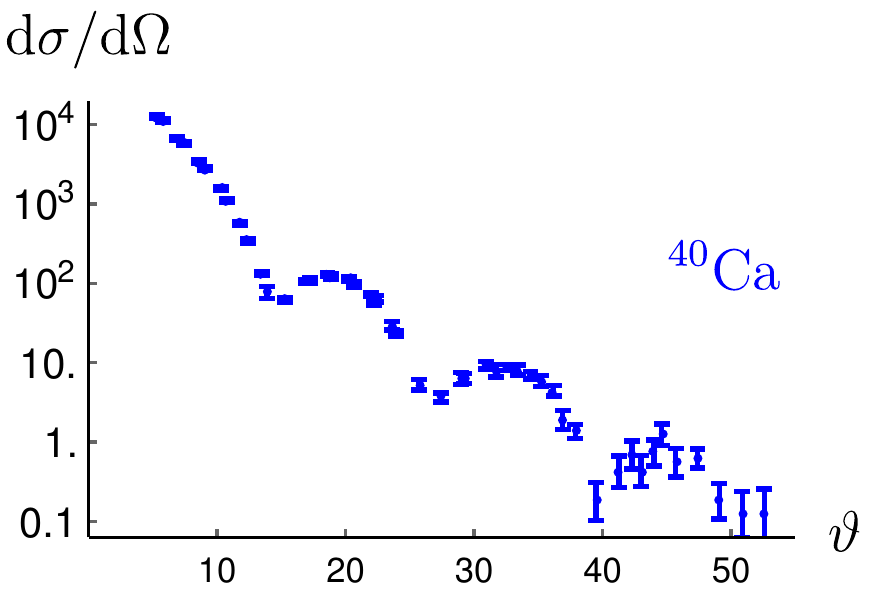}\quad
\includegraphics[width=.35\textwidth]{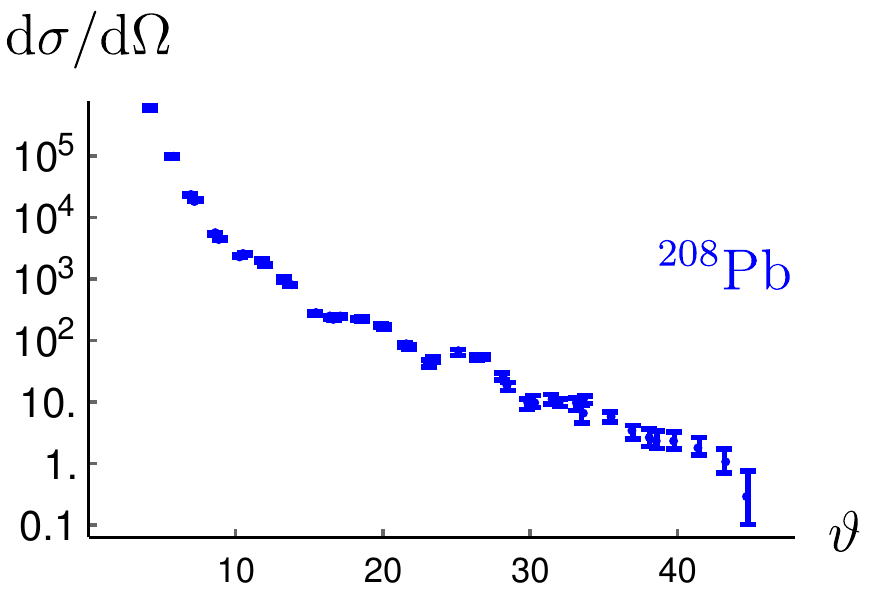}}
% antip-12C-179.pdf: 0x0 pixel, 300dpi, 0.00x0.00 cm, bb=
 \caption{Angular distribution for $\bar p$ scattering on the nuclei \isotope[12]{C}, \isotope[40]{Ca} and \isotope[208]{Pb} at kinetic energy $T_{\bar p}=180\,$MeV \cite{Garreta:1984rs}. The electronic retrieving of the data is due to Matteo Vorabbi. }
 \label{fig:PS184}
\end{figure}

At the start of the LEAR facility, the angular distribution of $\bar p\,\isotope[12]{C}$, $\bar p\,\isotope[40]{Ca}$ and $\bar p\,\isotope[208]{Pb}$ elastic scattering have been measured~\cite{Garreta:1984rs}.  Some of the  results are shown in Fig.~\ref{fig:PS184}. Data have been collected later at other energies and with other targets. %

\begin{figure}[ht!]
 \centerline{
 \includegraphics[width=.4\textwidth]{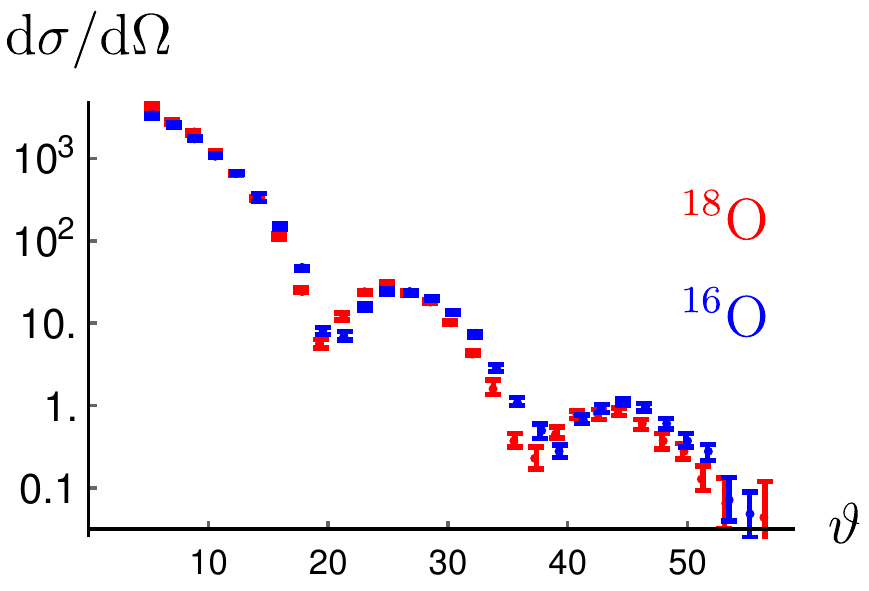}
 \qquad
 \includegraphics[width=.4\textwidth]{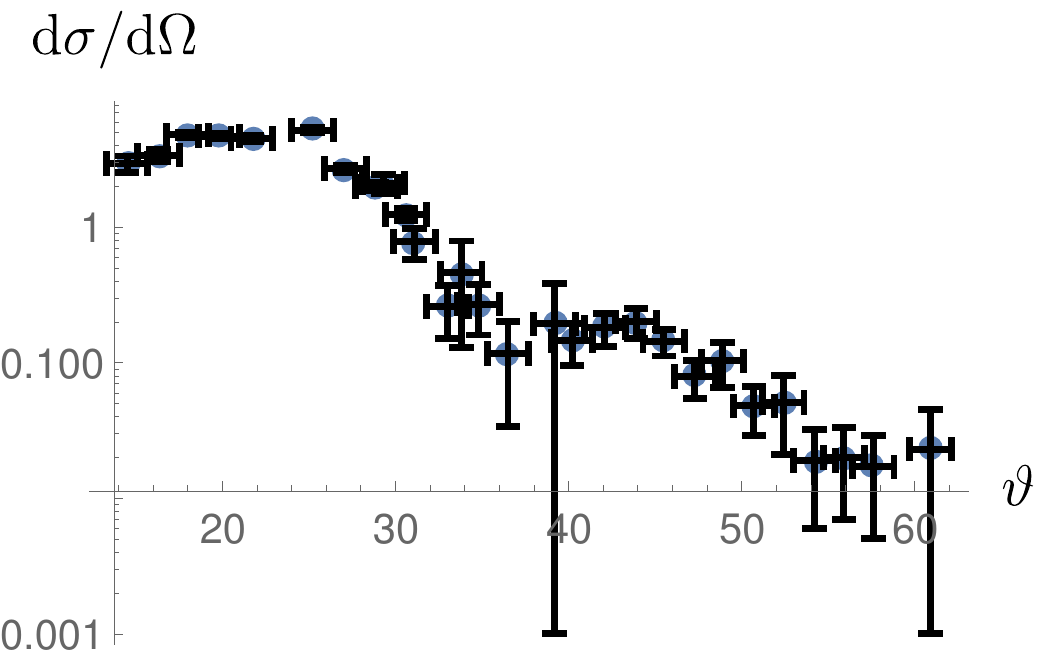}
 }
% antip-16O-178.pdf: 0x0 pixel, 300dpi, 0.00x0.00 cm, bb=
 \caption{Left: angular distribution for $\bar p$ scattering on \isotope[18]{O} and \isotope[16]{O} at 178.7\,MeV\cite{Bruge:1986fd}. Right: angular distribution of the $\isotope[12]C(\bar p,\bar p)\isotope[12]C^*$ reaction for the $3^-$ excited state at 9.6\,MeV.  The incident $\bar p$ has an energy of 179.7\,MeV}
 \label{fig:PS184a}
\end{figure}
The results have been interpreted in terms of an optical potential, either empirical, or 
 derived by folding the elementary $\bar N N$ amplitudes with the nuclear density, see, e.g., \cite{Adachi:1987zs,Suzuki:1985xv,Heiselberg:1986vi}.

 A comparison of \isotope[16]{O} and \isotope[18]{O} isotopes, see Fig.~\ref{fig:PS184a}, does not indicate any striking  isospin dependence of the $\bar pN$ interaction, when averaged on spins. However, at very low energy, some isospin dependence is suggested by the data and analyses  by the PS179 and OBELIX (PS201)\cite{Balestra:1988au,Botta:2001fu}.
\subsection{Inelastic scattering}
In between elastic scattering and annihilation, there is an interesting window of inelastic scattering, $\bar p A \to \bar p A^*$, where $A$ denotes an excited state of the nucleus $A$. It gives access to the spin-isospin dependence of the elementary $\bar N N$ amplitude, as stressed in \cite{Dover:1984cq}. A generalisation is the charge-exchange reaction $\bar p A \to \bar n B^{(*)}$.

Some results have been obtained by the PS184 collaboration on $\isotope[12]{C}$ and $\isotope[18]{O}$ \cite{Janouin:1986vh}, and analyzed in  in \cite{Dover:1984cq,Dover:1984yy}. See Fig.~\ref{fig:PS184a}. 
\subsection{Neutron-antineutron in nuclei}
The antinucleon-nucleon interaction plays a crucial role when estimating the lifetime of nuclei from  a given rate of neutron-to-antineutron transitions, as predicted in some theories of grand unification. The instability of nuclei is an efficient alternative to  free-neutron experiments, such as the one performed at Grenoble \cite{BaldoCeolin:1994jz}, with a limit of about $\tau_{n\bar n} \gtrsim 10^{-8}\,$s for the oscillation period.

A crucial feature is that the neutron-to-antineutron transition and the subsequent annihilation occurs at the nuclear surface. This alleviates the fear  \cite{Kabir:1983qx} that the phenomenon could be obscured in nuclei by uncontrolled medium corrections. Another consequence is that the schematic modelisation of nucleons and antineutrons evolving in a box, feeling an average potential $\langle V_n\rangle$ or $\langle V_{\bar n}\rangle$, does not work too well. It is important to account for the  tail of the neutron distribution, where $n$ and $\bar n$ are almost free. 

In pratice, there are several variants, see, e.g.,  \cite{Alberico:1998qy}. The simplest is based on the Sternheimer equation, which gives the first order correction to the wave function without summing over unperturbed states.  In a shell model with realistic neutron (reduced) radial  wave functions $u_{n\ell J}(r)$ with shell energy $E_{n\ell J}$, the induced $\bar n$ component is given by
\begin{equation}
 -\frac{w''_{n\ell J}(r)}{\mu}+\frac{\ell(\ell+1)}{\mu\, r^2}+ V_{\bar n}(r)\,w'_{n\ell J}(r) - E_{n\ell J}\,w'_{n\ell J}(r)= \gamma\,u_{n\ell J}(r)~.
\end{equation}
Here $\mu$ the reduced mass of the $\bar n$-$(A-1)$ system, $V_{\bar n}$ the complex (optical) $\bar n$-$(A-1)$ potential, and $\gamma=1/\tau_{n\bar n}$ the strength of the transition. Once $w_{n\ell J}$ is calculated, one can estimate the second-order correction to the energy, and in particular the width $\Gamma_{n\ell J}$ of this shell is given by 
\begin{equation}\label{eq:deut2}
\Gamma_{n\ell J}=-2\,\int_0^\infty \Im{V_{\bar n}}\,|w_{n\ell J}(r)|^2\,\mathrm{d}r=-2\,\gamma \int_0^\infty u_{n\ell J}(r)\, \Im{w_{n\ell J}(r)}\,\mathrm{d}r~,
\end{equation}
and is readily seen to scale as
\begin{equation}\label{eq:deut3}
 \Gamma_{n\ell J}\propto \gamma^2~.
\end{equation}
An averaging over the shells give a width per neutron $\Gamma$ associated with a lifetime~$T$
\begin{equation}\label{eq:deut4}
T=T_r\,\tau_{n\bar n}^2~,
\end{equation}
where $T_r$ is named either  the ``reduced lifetime'' (in s$^{-1}$) or the ``nuclear suppression factor''. The spatial distribution of the $w_{n\ell J}$ and the integrands in \eqref{eq:deut2}, the relative contribution to $\Gamma$ clearly indicate the peripheral character of the process. See, e.g., \cite{Barrow:2019viz,Barrow:2021svu} for an  application to a simulation in the forthcoming DUNE experiment and to Super-Kamiokande, and refs.\ there to earlier estimates. 

For the deuteron, an early calculation by Dover et al.~\cite{Dover:1982wv} gave $T_r\simeq 2.5\times 10^{22}\,\mathrm{s}^{-1}$. Oosterhof et al.~\cite{Oosterhof:2019dlo}, in an approach based on effective chiral theory, found a value significantly smaller, $T_r\simeq 1.1\times 10^{22}\,\mathrm{s}^{-1}$. However, their calculation has been revisited by Haidenbauer and Mei\ss ner \cite{Haidenbauer:2019fyd}, who got almost perfect agreement with Dover et al.  For \isotope[40]{Ar} relevant to the DUNE experiment, the result of  \cite{Barrow:2019viz} is $T_r\simeq 5.6\times 10^{22}\,\mathrm{s}^{-1}$.

\section{Antiprotonic atoms}
\subsection{Exotic atoms}
Exotic atoms are systems in which an electron is replaced by an heavier negatively charged particle. Among them, the muonic atoms $\mu^-p$ or $\mu^-A$, where $A$ denotes a nucleus, are useful tools to probe the structure of the positive kernel. However, it should be noted that the stability does not always survive the substitution $e^-\to\mu^-$. For instance, $p\mu^- e^-$ is unstable, unlike $\mathrm{H}^-(pe^-e^-)$.
Among exotic atoms, hadronic atoms $h^-p$ or $h^-A$, where $h^-$ denotes $\pi^-$, $K^-$ or $\bar p$, are of special interest, as they probe the interplay between the long-range Coulomb interaction and the short-range strong interaction. The simplest model consists of the Schr\"odinger equation
\begin{equation}\label{eq:Schr}
\left[-\Delta/(2\,\mu) +V_c+V_s-E\right]\Psi=0~,
 \end{equation}
 where $\mu$ is the reduced mass, $V_c$ is the Coulomb term, $V_s$ the $h^-$-nucleus strong-interaction potential, and $E$ the energy, whose difference $\delta E=E-E_c$ with respect to the pure Coulomb case $E_c$ is referred to as the level shift. In most cases, ordinary perturbation theory is not suited to estimate $\delta E$ and the deformation of the wave function $\Psi$. For instance,  a hard core of small radius gives a tiny $\delta E$, while the first order correction is infinite! The expansion scheme is actually the ratio $a/R$ of the scattering length in $a_s$ to the Bohr radius. For $S$ waves, the leading term is 
 \begin{equation}\label{eq:DT}
  \delta E\simeq -E_c\,\frac{4\,a}{n\,R}~,
 \end{equation}
 where $n$ is the principal quantum number. This formula, referred to as the Deser-Trueman formula~\cite{Deser:1954vq,Trueman:1961zza}, works quite well, as long as $|a|\ll R$. It can be extended to non-Coulomb interaction in the long range, see, e.g., \cite{2011JPhA...44A5302C}. More recently, the problem has been formulated in the framework of effective theories \cite{Holstein:1999nq}.
 \subsection{Level rearrangement}
 The changes undergone by \eqref{eq:DT} at large $|a|$ are better seen the way initiated by Zel'ldovich \cite{Zeldo:1960} and Shapiro et al.~\cite{Shapiro:1978wi}, who rewrote \eqref{eq:Schr} with a potential $V_c+\lambda\,V_s$, and studied how the spectrum evolves as $\lambda$ is varied. 
A typical example is shown in Fig.~\ref{fig:rearr}. If the range of $V_s$ is made shorter, then the behavior is sharper,  almost like a staircase function. See, e.g., \cite{Deloff:2003ns,Combescure:2007ki}, and also \cite{Gal:1996pr}.
\begin{figure}[ht!]
 \centering
 \includegraphics[width=.5\textwidth]{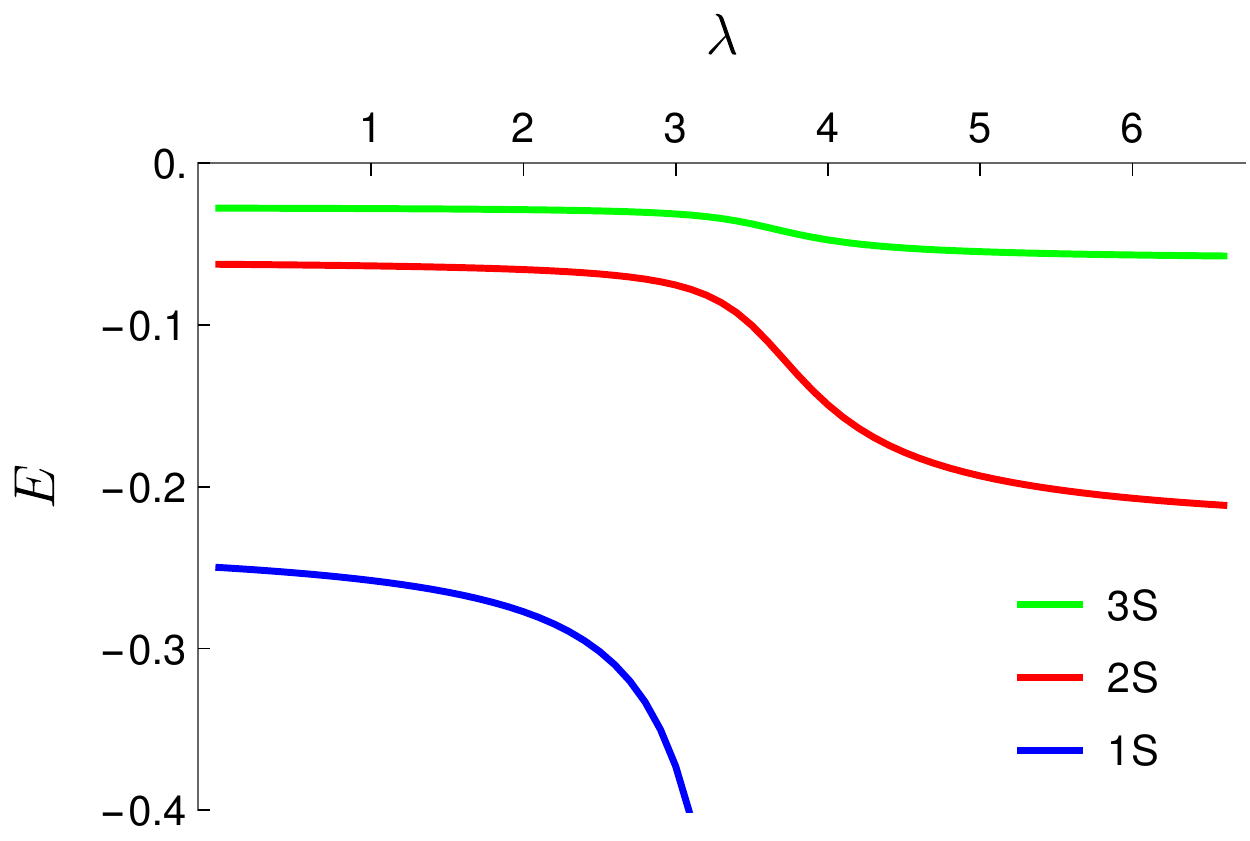}
 \caption{First three levels of the S-wave solution of \eqref{eq:Schr}  with $V(r)=-1/r+\lambda\,a^2 \exp(-a\,r)$. The pattern of rearrangement is shown here for $2\,\mu=1$ in \eqref{eq:Schr}, and $a=100$ in the potential $V(r)$.}
 \label{fig:proto}
\end{figure}
As a consequence, an anomalously large energy shift would indicate a bound or virtual state near the threshold. 
\subsection{Protonium}
The analysis of protonium is more delicate than the above one-channel modeling. At long distance, the wave function is pure $\bar p p$. At shorter distances, there is an admixture of $\bar n n$, and one of the isospin components, $I=0$ or $I=1$, dominates, depending on the partial wave. This is important to understand the branching ratios of protonium annihilation. 

For angular momentum $\ell>1$, the energy shifts are negligible. For $\ell=1$, they are dominated by the one-pion-exchange tail of the interaction, as shown in \cite{Kaufmann:1978vp}. For $l=0$, there is a net effect of the absorptive part of the interaction, with $\delta E$ complex, and a positive real part, i.e., a repulsive effect. 

For the experimental results, see, e.g., \cite{AUGSBURGER1999149} and refs.\ therein. The average shift is $\delta E= (712.5\pm 20.3)-i\,(527\pm 33)\,$eV. An attempt to separate  $\isotope[1]{S}_0$ and $\isotope[3]{S}_1$ gives a smaller shift and a larger width for the former, in agreement with the potential-model calculations \cite{Richard:1982zr,Carbonell:1992wd,Gutsche:1998fc}.
\subsection{Antiproton-nucleus atoms}
The latest calculation of the antiprotonic-deuterium atom has been performed in~\cite{Lazauskas:2021cdi}, with an optical potential input into the Faddeev equations. References can be found there to earlier estimates based on mere folding of the $\overl N N$ amplitude with the deuterium wave function. 

For heavier nuclei, there are fits based on an empirical Wood-Saxon potential or on a potential proportional to the nuclear density $\varrho(r)$ \cite{Batty:1997zp,Friedman:2007zza}
\begin{equation}
 U(r)=-\frac{2\,\pi}{\mu}\left(1+\frac{\mu}{m}\right)\,\bar a\,\varrho(r)~,
\end{equation}
where $m$ is the antiproton mass, $\mu$ the reduced mass, and $\bar a$ an effective (complex) scattering lenght. A good fit is obtained with $\Re{\bar a}\sim \Im{\bar a}\sim 1\,$fm.  More ambitious are the attempts to derive the potential from the ``elementary'' $\overl N N$ interaction \cite{Dumbrajs:1986bh,Adachi:1987zs,Suzuki:1983rv}.
\subsection{Antiprotonic helium}
In 1964, Condo proposed that metastable states could be formed in exotic Helium atoms $\mathrm{He}\,h^-\,e^-$ \cite{1964PhL.....9...65C,Russell:1969zz}. Such states have been observed at CERN in antiprotonic Helium \cite{PS205:1993ypr}. At first, such states look just as a curiosity of the three-body problem. Actually, antiprotonic atoms have emerged as a remarkable precision laboratory to measure the antiproton properties (mass, charge, \dots) and even the fine-structure constant~\cite{Yamazaki:2002he}. 
\subsection{Day-Snow-Sucher effect}
The process of formation of antiprotonic atoms has been much studied \cite{Borie:1980wol}. A low-energy antiproton is slowed down by electromagnetic interactions and captured in some high orbit. The electrons of the initial atoms are expelled during this capture and the series of transitions of the antiproton towards lower states, preferentially through circular orbits with $\ell=n-1$. In a vacuum,
protonium annihilation is negligible in high orbits with $l>1$, at the level of about $1\,\%$ in 2P, and $100\,\%$ in 1S~\cite{Kaufmann:1978vp}. However, in a dense target, it often happens that the protonium state goes inside the ordinary atoms, where it experiences a Stark mixing of levels with the same principal number $n$ but orbital momentum lower than $\ell=n-1$, eventually down to $\ell=0$, where annihilation might occur. This is the Day-Snow-Sucher effect \cite{Day:1960zz}. It plays a crucial role when analyzing the results of annihilation at rest: the changes of the  branching ratios as a function of  the density of the target  allow one to determine the contributions of S-waves vs.~the ones of P-waves. 
\section{Antiprotons in the Universe}
A major issue in cosmology is why matter is seemingly dominant over antimatter, and whether some pieces of antimatter remain. For instance, the AMS experiment,installed on the International Space Station, has measured a $\bar p/p$ ratio of about $0.2\times 10^{-3}$~\cite{AMS:2016oqu}. AMS has also detected antinuclei in cosmic rays and one should discuss whether they are primary or secondary objects.

The question of antimatter in the Universe is recurrently addressed. Till the 70s, some models were elaborated with a symmetric (matter vs.\ antimatter) Universe, with the key issue of how matter domains have been separated from the antimatter ones. In this context, an amusing correlation was made between a positive energy  shift in protonium and a repulsive interaction (due to annihilation) between matter and antimatter~\cite{Caser:1972cb}. The more recent scenarios assume that thanks to $CP$ violation (charge conjugation times parity)~\cite{Sakharov:1967dj}, the Universe is dominated by matter from its very beginning. Now a very small amount of $CP$ violation has been detected in neutral kaons \cite{Christenson:1964fg} and later in mesons carrying heavy flavor \cite{bigi2000cp}, but this is just of beginning.  
\section{Outlook}
The physics of low-energy antiprotons is extremely rich at the interface of nuclear physics and quark physics, and also atomic physics and physics beyond the standard model. The experimental activity has been dramatically vitalized in the 80s with the advent of stochastic cooling. Today, a very ambitious program of high-precision measurements is carried out with very low energy antiprotons. In the near future, the physics of charm and strangeness should get interesting information from medium-energy antiprotons. 

\vskip .2cm
\noindent
\textbf{Acknowledgments:} The author thanks M.~Asghar for useful comments. 
%\bibliographystyle{aipauth4-1}
%\bibliographystyle{plainnat}
%\bibliographystyle{apsnat}
%  \bibliographystyle{unsrt}
%  \bibliography{nnb1}

%
\end{document}